\documentclass{ieeeaccess}
\usepackage{cite}
\usepackage{amsmath,amssymb,amsfonts}
\usepackage{amssymb}
\usepackage[export]{adjustbox}
\usepackage{array}
\usepackage[utf8]{inputenc}
\usepackage{booktabs}
\usepackage{enumerate}
\usepackage{esint}
\usepackage{tabularx}
\usepackage[T2A,LGR,T1]{fontenc}
\usepackage{multirow}
\usepackage{graphicx}
\usepackage{textcomp}
\def\BibTeX{{\rm B\kern-.05em{\sc i\kern-.025em b}\kern-.08em
    T\kern-.1667em\lower.7ex\hbox{E}\kern-.125emX}}
\begin{document}
\history{Date of publication xxxx 00, 0000, date of current version xxxx 00, 0000.}
\doi{10.1109/ACCESS.2023.0322000}

\title{Decentralized Federated Learning on the Edge over Wireless Mesh Networks}
\author{\uppercase{Abdelaziz Salama}\authorrefmark{1},\IEEEmembership{Member, IEEE}, \uppercase{Achilleas Stergioulis}\authorrefmark{1}, \IEEEmembership{Member, IEEE},
Syed Ali Zaidi\authorrefmark{1},\IEEEmembership{Member, IEEE}, and Des McLernon\authorrefmark{1}
\IEEEmembership{Member, IEEE}}

\address[1]{Department of Electrical and Electronic Engineering, University of Leeds, Leeds, UK}

\tfootnote{This research was funded by EP/X040518/1 EPSRC CHEDDAR and was partly funded by UKRI Grant EP/X039161/1 and MSCA Horizon EU Grant 101086218.}

\markboth
{Abdelaziz \headeretal: Preparation of Papers for IEEE TRANSACTIONS and JOURNALS}
{Abdelaziz \headeretal: Preparation of Papers for IEEE TRANSACTIONS and JOURNALS}

\corresp{Corresponding author: Abdelaziz Salama (elamasa@leeds.ac.uk)}

\begin{abstract}
The rapid growth of Internet of Things (IoT) devices has generated vast amounts of data, leading to the emergence of federated learning as a novel distributed machine learning paradigm. Federated learning enables model training at the edge, leveraging the processing capacity of edge devices while preserving privacy and mitigating data transfer bottlenecks. However, the conventional centralized federated learning architecture suffers from a single point of failure and susceptibility to malicious attacks. In this study, we delve into an alternative approach called decentralized federated learning (DFL) conducted over a wireless mesh network as the communication backbone. We perform a comprehensive network performance analysis using stochastic geometry theory and physical interference models, offering fresh insights into the convergence analysis of DFL. Additionally, we conduct system simulations to assess the proposed decentralized architecture under various network parameters and different aggregator methods such as FedAvg, Krum and Median methods. Our model is trained on the widely recognized EMNIST dataset for benchmarking handwritten digit classification. To minimize the model's size at the edge and reduce communication overhead, we employ a cutting-edge compression technique based on genetic algorithms. Our simulation results reveal that the compressed decentralized architecture achieves performance comparable to the baseline centralized architecture and traditional DFL in terms of accuracy and average loss for our classification task. Moreover, it significantly reduces the size of shared models over the wireless channel by compressing participants' local model sizes to nearly half of their original size compared to the baselines, effectively reducing complexity and communication overhead.
\end{abstract}

\begin{keywords}
Internet of Things (IoT),
Federated Learning,
Decentralized Federated Learning,
Edge Computing and 
Data Privacy.
\end{keywords}

\titlepgskip=-21pt
\maketitle

\section{Introduction}
\label{sec:introduction}
\PARstart{I}{n} recent years, the proliferation of Internet of Things (IoT) devices has been remarkable, largely driven by advancements in 5G technology and beyond. The global IoT market is projected to grow to a market value of 1.567 trillion US dollars by 2025 \cite{vailshery_2021}. The architecture of IoT, which incorporates spectrum and energy management mechanisms, is designed to optimize both spectral and energy efficiencies to their maximum potential  \cite{afzal2015cognitive}. Alongside this, breakthroughs in semiconductor technology have enabled the fabrication of transistors with sub-10 nanometer gate lengths, drastically improving processing capacity while reducing power consumption \cite{ShalfJohnM2015CbML}. These developments play a crucial role in the extensive adoption of embedded devices, including smartphones, sensors and tablets. These devices require powerful microprocessors capable of delivering high performance while adhering to stringent energy consumption limitations. 

\begin{figure}[t]
\includegraphics[scale=0.35]{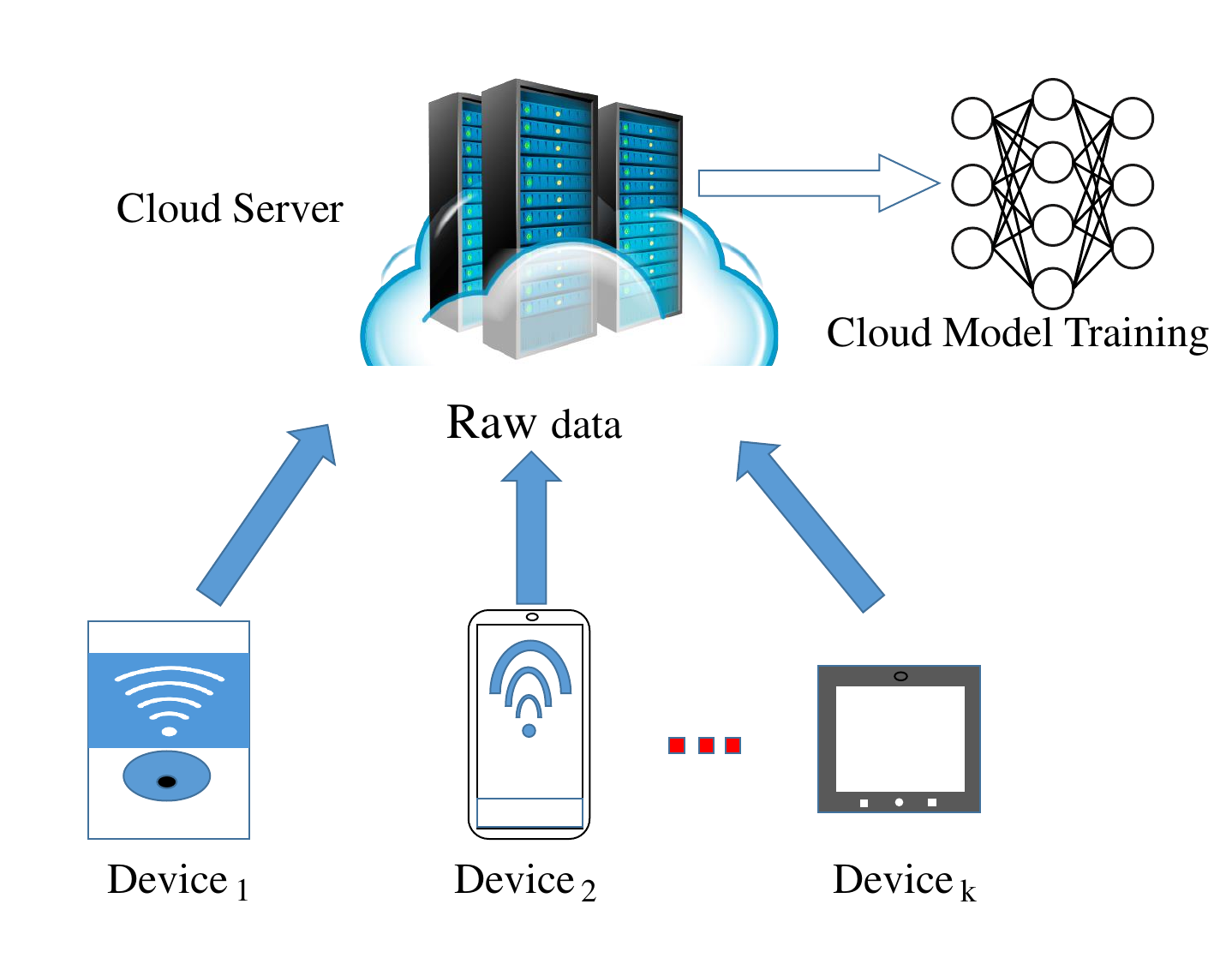}
\caption{Typical data flow and analysis in IoT systems.}
\label{fig1}
\end{figure}

The substantial volume of data generated by embedded sensor devices at the periphery of IoT systems has given rise to the field of Big Data, which focuses on devising effective methods for processing, disseminating, and analyzing extensive datasets. In typical IoT setups, data initially collected by sensors at the edge are transmitted through a central network to a cloud server as shown in Figure (\ref{fig1}), where various data preprocessing techniques (such as data cleansing, feature extraction, denoising, etc.) are applied. Subsequently, this processed data is employed for training machine learning models tailored to the specific application domain \cite{iot_Data}. These trained models serve various machine learning tasks, including classification, clustering, anomaly detection, and regression, facilitating precise and optimal decision-making. The data flow structure described above in IoT systems has benefited from advancements in High-Performance Computing (HPC) systems, enabling real-time and low-latency solutions across diverse industry sectors.

Despite its widespread adoption, the conventional centralized cloud processing architecture has several limitations when considering practical scenarios. 

Firstly, the reliance on transferring a large volume of data from edge devices to the cloud for analysis introduces significant challenges related to communication channel capacity and system latency. This becomes particularly problematic for IoT systems that utilize low-power, low-data-rate communication protocols such as Zigbee and LoRa \cite{lora}. 

Secondly, the nature of the data collected by embedded sensor devices often involves privacy-sensitive information, making the transmission of raw data over potentially insecure network connections a significant concern. Data breaches can have severe consequences for the data owners and are strictly regulated by international legislation such as the European General Data Protection Regulation (GDPR) \cite{gdpr}.

Furthermore, the increased processing capacity of edge devices remains underutilized as they are primarily used for data gathering and transmission purposes, neglecting their potential for local computation and analysis.

\subsection{Motivation}
\begin{figure}[t]
\centerline{\includegraphics[scale=0.37]{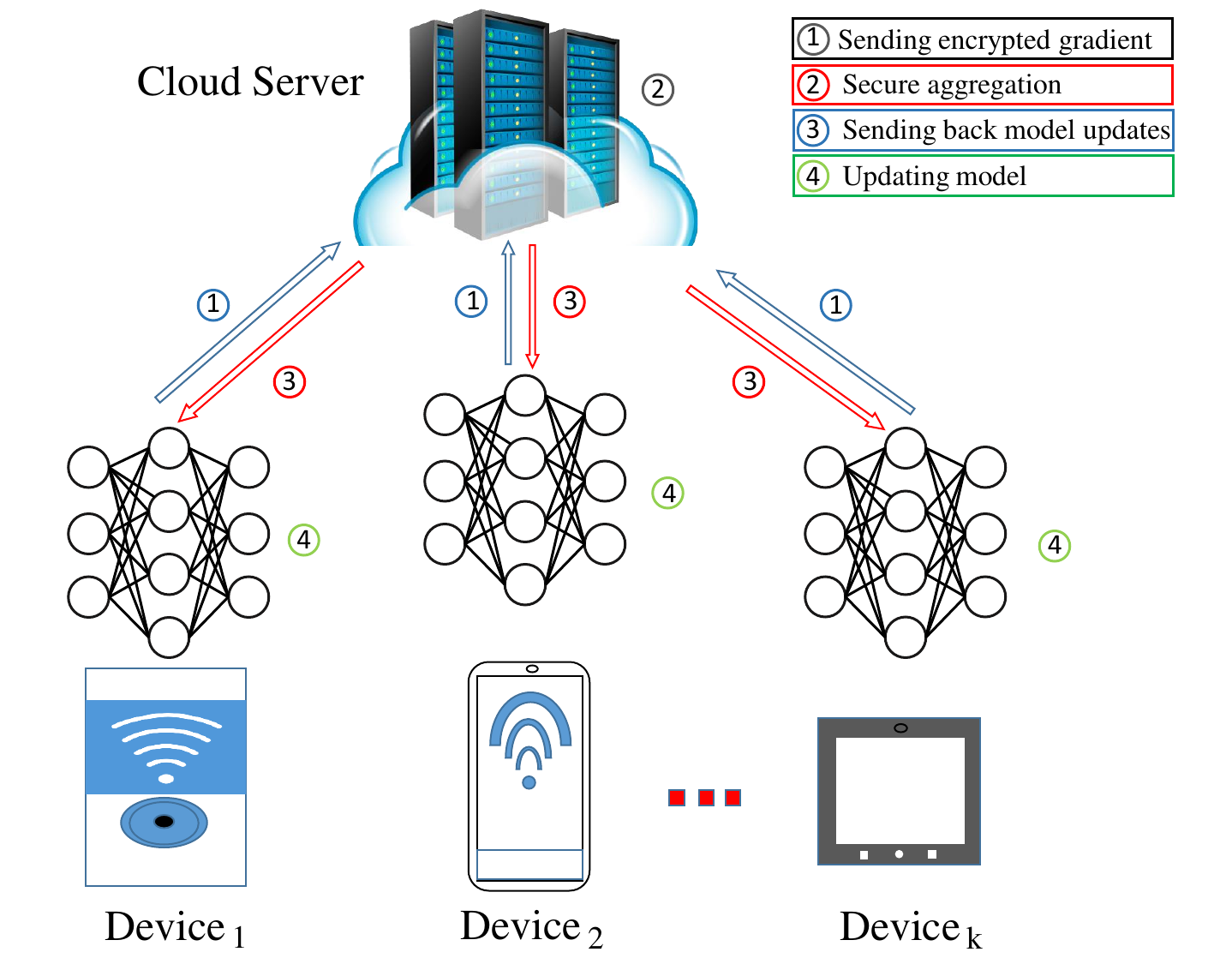}}
\caption{Centralized FL system architecture.}
\label{fig2}
\end{figure}

In 2017, Google researchers introduced an enhancement to the traditional centralized data flow architecture by proposing Federated Learning (FL) as an innovative distributed model learning framework that operates across the entire system \cite{mcmahan2017communicationefficient, kone2017federated}. FL revolutionizes the model training process by ensuring that raw data remains localized on the device where it is generated, thereby preserving privacy and enhancing data security. In this approach, a global model is periodically updated by a centralized server in the cloud and shared with edge devices during each training iteration. Each device independently refines the global model using its local dataset and transmits only the updated model parameters back to the cloud server. The server then aggregates these model parameter updates from all participating devices to perform a global model update. This iterative process continues until the global model converges based on a predefined threshold value. Figure \ref{fig2} illustrates the structure of a typical FL system with a central coordinating server.

Although the initial FL architecture represents a substantial advancement, particularly in terms of privacy and overall system efficiency, it has notable limitations, primarily stemming from its centralized design. One significant drawback is the central server, which serves as a singular point of failure, compromising the system's resilience. This vulnerability is especially concerning for time-sensitive IoT applications, as any temporary server downtime can lead to severe disruptions. Additionally, the centralization of data, even if limited to transmitting model updates, escalates the risk of malicious attacks \cite{privacy}.

\begin{figure}[t]
\centerline{\includegraphics[scale=0.37]{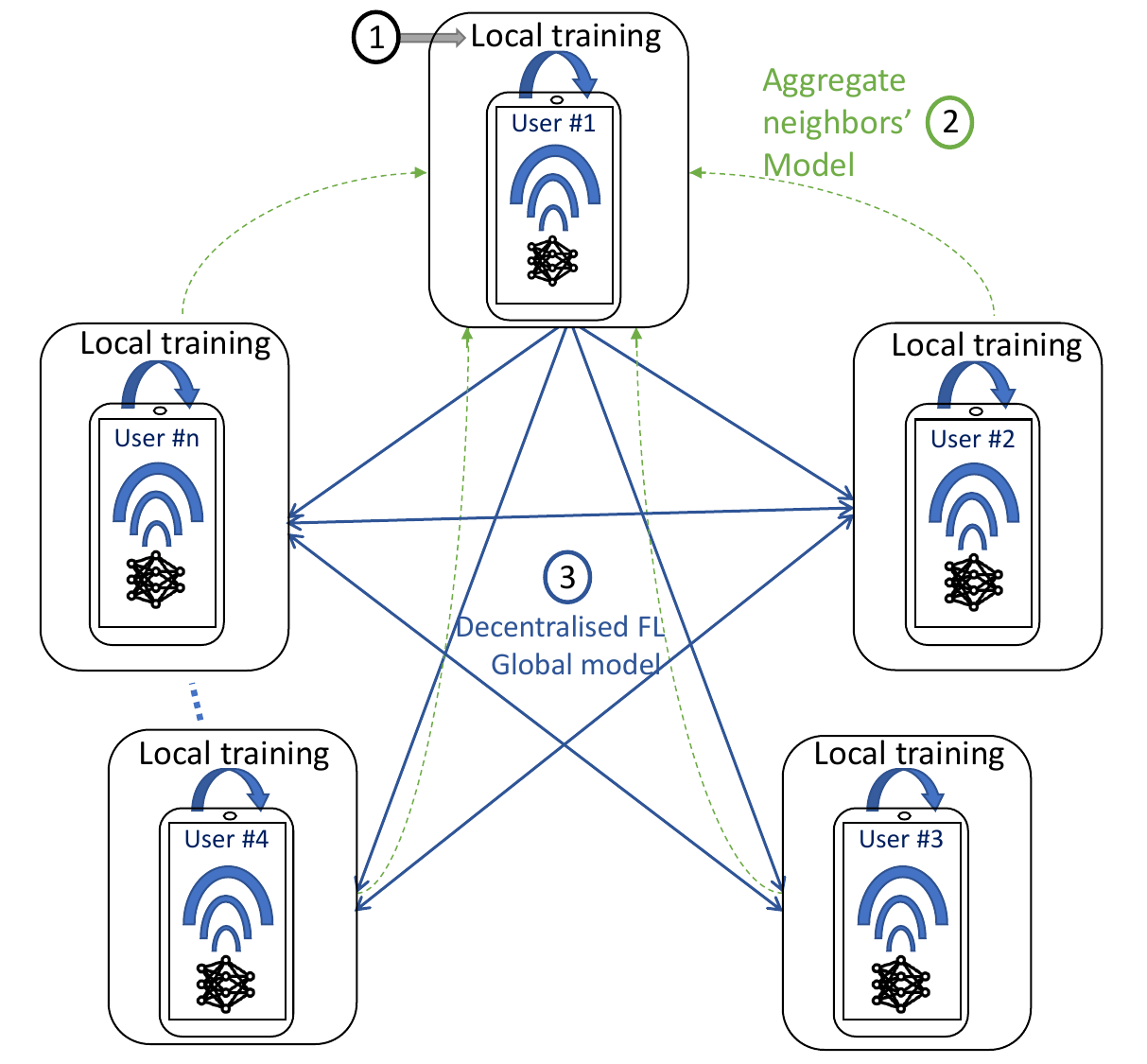}}
\caption{Decentralized FL system architecture.}
\label{fig3}
\end{figure}

In this work, a fully decentralized FL architecture is proposed and analysed to overcome the centralized FL limitations. In our system, there exists no central coordinating entity, such as a cloud server, and the model learning process is fully distributed among network participants, namely the edge devices. More specifically, each individual edge device is responsible for maintaining and updating its local model. Unlike the traditional centralized FL architecture, there is no overarching global model in our approach. During each training iteration, every device shares its current local model parameters with neighbouring devices and receives their respective updates in return. Subsequently, it performs a local model update using its unique dataset. The device then aggregates the received model parameters with its own, calculating the local model parameters for the next iteration. This decentralized learning architecture has been empirically demonstrated to converge to an equivalent central model through the application of gossip training theory in distributed network learning \cite{kempe} and \cite{diffusion}, as illustrated in Figure \ref{fig3}.

In the realm of Decentralized Federated Learning (DFL) on edge devices within wireless mesh networks (WMNs), we believe that our paper makes significant contributions on several fronts. Our primary focus is on reducing communication overhead through the application of cutting-edge compression techniques. Finally, we believe this work makes significant contributions in the area of Federated Learning and network performance analysis.

\subsection{Key contribution}
In this paper, we present a comprehensive description of the DFL architecture within the context of the Internet of Things (IoT) and edge devices while emphasizing the significance of the underlying core communication network. Our unique approach is distinct from much of the existing literature, as we aim to provide an all-encompassing perspective on the decentralized FL paradigm while considering the performance implications of the communication network that facilitates data transfer among IoT edge devices. We have chosen wireless one-hop mesh networking as the foundational network infrastructure, aligning with the distributed nature of IoT network learning. The rationale behind this choice is twofold.

Firstly, mesh networking possesses the adaptive capability to reconfigure their routing graph in real-time, particularly when faced with link interruptions \cite{AKYILDIZ2005445, benyamina}. This adaptive characteristic proves invaluable in real-world scenarios where edge devices are often deployed in challenging environments, prone to link failures. By harnessing the dynamic nature of the mesh networks, we ensure consistent and reliable data transfer, even in the presence of intermittent connectivity or link disruptions. This resilience is critical to maintaining the integrity of the decentralized FL process, enabling the system to gracefully handle disruptions without compromising learning.

\begin{table*}
\caption{An assessment of surveys exploring CFL and DFL studies}.
\label{table:research-summary}
\centering
\begin{tabular}{|l|l|l|l|l|l|l|l|}
\hline
\textbf{Ref.} & \textbf{Year} & \begin{tabular}[c]{@{}l@{}}\textbf{FL} \\ \textbf{Approach} \end{tabular} & \begin{tabular}[c]{@{}l@{}}\textbf{Device Types} / \\ \textbf{Area} \end{tabular} & \begin{tabular}[c]{@{}l@{}}\textbf{Various} \\ \textbf{Aggregators} \end{tabular} & \begin{tabular}[c]{@{}l@{}}\textbf{Frame-} \\ \textbf{works} \end{tabular}  & \begin{tabular}[c]{@{}l@{}}\textbf{Appli-} \\ \textbf{cation} \end{tabular}  & \textbf{Focus and Solution Categorization} \\ \hline
\cite{lim2020federated} & 2020 & CFL & Mobile networks & \checkmark & \checkmark & \sffamily x & \begin{tabular}[c]{@{}l@{}}A survey on FL and edge computing in mobile \\networks.\end{tabular} \\ \hline
\cite{nguyen2021federated} & 2021 & CFL & IoT devices & \sffamily x & \sffamily x & \checkmark & \begin{tabular}[c]{@{}l@{}}A review of FL in the context of IoT, data \\distribution, privacy and ML models architectural. \end{tabular} \\ \hline
\cite{khan2021federated} & 2021 & CFL & IoT devices & \checkmark & \sffamily x & \checkmark & \begin{tabular}[c]{@{}l@{}}A systematic review of FL in IoT, detailing \\limitations and applications.\end{tabular} \\ \hline
\cite{mothukuri2021survey} & 2021 & CFL & Security Privacy & \sffamily x & \checkmark & \sffamily x & \begin{tabular}[c]{@{}l@{}}A survey of FL security and privacy, defining \\secure protocols.\end{tabular} \\ \hline
\cite{boobalan2022fusion} & 2022 & CFL & FL baselines & \checkmark & \checkmark & ? & \begin{tabular}[c]{@{}l@{}}A survey on the FL baselines with a basic \\introduction to definitions and architectures.\end{tabular} \\ \hline
\cite{joshi2022federated}& 2022 & CFL & Healthcare & \checkmark & \sffamily x & \checkmark & \begin{tabular}[c]{@{}l@{}}An overview of the implement of FL as a tool \\in the healthcare scenarios.\end{tabular} \\ \hline
\cite{witt2022decentral} & 2022 & CFL & \begin{tabular}[c]{@{}l@{}}Framework \\review \end{tabular}& \checkmark & \checkmark & ? & \begin{tabular}[c]{@{}l@{}}An overview of frameworks for deploying \\DFL-based architectures.\end{tabular} \\ \hline
\cite{qu2022blockchain} & 2022 & DFL & DLT & \sffamily x & \sffamily x & ? & \begin{tabular}[c]{@{}l@{}}A description of the challenges and applications \\of Blockchain.\end{tabular} \\ \hline
\cite{billah2022systematic} & 2022 & DFL & IoV & \sffamily x & \checkmark & ? & \begin{tabular}[c]{@{}l@{}}A brief description of FL in fog radio access \\networks.\end{tabular} \\ \hline
\cite{gupta2022survey} & 2022 & DFL & \begin{tabular}[c]{@{}l@{}}Wireless \\communications \end{tabular} & \sffamily x & \sffamily x & \checkmark & \begin{tabular}[c]{@{}l@{}}A review of the techniques for adapting \\FL to distributed environments.\end{tabular} \\ \hline
\cite{saraswat2022blockchain} & 2022 & DFL & UAV devices & \sffamily x & \sffamily x & \checkmark & \begin{tabular}[c]{@{}l@{}}A brief survey on the application of FL in UAV \\networks.\end{tabular} \\ \hline
\cite{wu2023topology} & 2023 & DFL & DFL baselines & \sffamily x & \sffamily x & \checkmark & \begin{tabular}[c]{@{}l@{}}A study of optimized DFL models and algorithms\\ focusing on network topologies.\end{tabular} \\ \hline
\cite{chen2023advancements} & 2023 & DFL & DFL baselines & \sffamily x & ? & \checkmark & \begin{tabular}[c]{@{}l@{}}A comparison of CFL and DFL federation \\architectures regarding topologies, privacy, and \\security.\end{tabular} \\ \hline
This work & 2023 & DFL & \begin{tabular}[c]{@{}l@{}}Edge devices \\and IoT devices \end{tabular}  & \checkmark & \checkmark & \checkmark & \begin{tabular}[c]{@{}l@{}} Analyzed CFL and DFL fundamentals, commun-\\ication management, geometric network analysis\\ and employed diverse aggregator methods.\end{tabular} 
\\ 
\hline
\end{tabular}
\end{table*}

This work combines the DFL architecture with wireless mesh networking, establishing a robust and efficient framework for collaborative learning within IoT systems. Our analysis delves beyond the FL architecture's performance, encompassing the unique attributes of the communication networks. We use one-hop communication mesh networking to ensure efficient data transfer among edge devices. Additionally, we employ different aggregator methods to update the global model and achieve optimum performance. This comprehensive perspective enhances our understanding of the overall system behaviour. Our research aims to contribute valuable insights into the design and optimisation of DFL systems in IoT environments, promising improved performance and scalability in real-world applications. Moreover, we harness the innate strengths of the mesh networks, such as resilience to single-point failures and heightened data privacy, fortifying the security and reliability of FL within IoT systems. This crucial aspect of our research addresses the mounting complexity and risk factors in large-scale distributed learning applications effectively.

To further enhance our model and minimize communication overhead, we have implemented a state-of-the-art compression technique, as exemplified by the genetic algorithm-based approach \cite{agarwal2023genetic}, to reduce the dimensions of terminal models at the edge. In the case of the Convolutional Neural Network (CNN) model, this reduction entails selecting a subset of convolutional filters and nodes within the dense layers while ensuring that the original models' accuracy levels remain intact.

\subsection{Organisation}
The rest of this report is organised as follows: In section \ref{sec: Theoretical Background}, background and related work from literature is presented, and a complete theoretical analysis of the performance of wireless mesh networks and the convergence criteria of the proposed DFL system architecture is carried out. Section \ref{sec: methodology} introduces the methodology for the learning process, network topology, communication methods, and the optimal approach for selecting highly reliable devices to update the global model that is updated using a compressed model at the edge and a range of aggregator methods. In section \ref{sec: Simulation and Experiments}, the simulation setup is described, and the corresponding results are presented along with their interpretation. Section \ref{sec: results and discussion} presents the results of the simulations and in-depth discussions, offering insights and analysis of the findings.
Finally, in section \ref{sec: Conclusion and Future Work} a summary of the work is provided, and future work directions and milestones are proposed.

\section{BACKGROUND AND RELATED WORK}
\label{sec: Theoretical Background}
Ever since its introduction in \cite{mcmahan2017communicationefficient}, Federated Learning (FL) has gained considerable attention and has become one of the most extensively researched machine learning paradigms. 

The literature surrounding FL is expansive, encompassing diverse architectures, analyses of learning performance, and investigations into data privacy concerns. While the majority of research initially gravitated towards the conventional centralized architecture of FL, recent efforts have increasingly shifted their focus towards decentralized alternatives called decentralized FL (DFL), where basically no central server is needed \cite{lee2022device}.

Related work on DFL can be distinguished into two main design philosophies. The primer approach to achieving decentralization involves harnessing blockchain technology, a highly promising avenue. In these blockchain-based systems, participants fall into two categories: standard edge devices and miners. Each edge device establishes communication with nearby miners, who assume the role of model aggregators during particular training rounds. Depending on the employed consensus algorithm, the miner that successfully solves the hashing problem earns the privilege of atomically updating the distributed ledger with the new global model update. In the study \cite{10049061}, the limitations of centralized federated learning (CFL) are addressed by proposing a decentralized federated learning (DFL) approach that eliminates the need for a central server and instead relies on one-hop neighbours for collaboration in the communication network. They use stochastic geometry to model the dynamics of the network topology, MAC protocol, and fading on links, allowing them to evaluate the performance of DFL while preserving privacy and accommodating networking dynamics. However, this study primarily focuses on the evaluation of DFL without considering its application on the edge and evaluating the network intensities for multi-hop wireless mesh networks. 

Furthermore, in our research efforts, we have undertaken an extensive assessment of relevant literature and surveys pertaining to both CFL and DFL. This comprehensive evaluation is meticulously presented in table (\ref{table:research-summary}), which delves into several facets of CFL and DFL, including global model aggregation methods, foundational frameworks, application domains, and categorizations based on proposed solutions. To facilitate easy interpretation, symbols have been employed within the table to signify the status of each aspect: a checkmark ($\checkmark$) indicates full coverage, a question mark (?) denotes partial coverage, and a multiplication symbol ($\text{\sffamily x}$) signifies that the particular aspect has not been addressed.

In \cite{lee2022device}, authors aimed at optimizing the overall average number of parameter transmissions only in the CFL approach, including shallow and complete transmissions, while maintaining a predefined ratio between them. To offer a thorough analysis of DFL within the context of core communication network performance, table (\ref{table1}) has been compiled to provide an overview of recent developments in CFL and DFL, which serves as an overview of recent developments in the field and their primary focus areas. This table provides a critical evaluation of contemporary advancements in FL network design, spanning multiple dimensions such as resource management, system cost, security, privacy, user distribution analysis, communication network characteristics, FL network intensity, performance, and central server-free approaches.

The works in \cite{balasubramanian2021intelligent} and \cite{mahmod2023improving} investigate the challenges posed by the widespread deployment of Internet of Things (IoT) devices in the 5G era, particularly in the context of software-defined networks (SDNs). It highlights the importance of cache management at the edge of the network and explores emerging edge resources like mobile device clouds and micro-edge data centres. The goal is to optimize content placement based on user demand and cost considerations. The study also addresses security and seamless data delivery in mobile IoT networks and introduces federated learning (FL) as a key framework to harness data and computational resources from end-user devices for training machine learning models. The paper's main focus is on centralized federated learning in the 5G network, leaving potential opportunities in decentralized learning methods, particularly in Ad-hoc networks, relatively unexplored.

\begin{table*}
\caption{Summary on FL-related topics with our paper's contribution.}
\label{table1}
\centering
\begin{tabular}{| >{\centering\arraybackslash}m{1.59cm}|p{1.59cm}|m{1.59cm}|m{1.59cm}|m{1.59cm}|m{1.59cm}|m{1.59cm}|m{1.59cm}|}

\hline

   \multicolumn{1}{|c|}{\multirow{2}{*}{Related Research}} &
  \multicolumn{1}{c|}{\multirow{2}{*}{Main research area}} &
  \multicolumn{6}{c|}{Assessment of recent developments in the design of FL networks} \\ \cline{3-8} 
\multicolumn{1}{|c|}{} &
  \multicolumn{1}{c|}{} &
  Allocation of resources and cost management &
  Privacy and security &
  Analyzing the distribution of users &
  The network communication &
  FL Network intensity and performance &
  Central Server-free\\ 
 \hline
\cite{ye2022decentralized} & DFL concept & \text{\sffamily x} & \checkmark & \text{\sffamily x} & \checkmark & \text{\sffamily x} & \checkmark \\ \hline
\cite{nguyen2021federated} & FL concept & \text{\sffamily x} & \checkmark & \text{\sffamily x} & \checkmark & \text{\sffamily x} & \checkmark \\ \hline
\cite{9670674} & Distributed ML & \checkmark & \text{\sffamily x} & \text{\sffamily x} & \checkmark & \text{\sffamily x} & \text{\sffamily x} \\ \hline
\cite{mothukuri2021survey} & Security and Privacy in FL & \text{\sffamily x} & \checkmark & \text{\sffamily x} & \text{\sffamily x} & \text{\sffamily x} & \checkmark \\ \hline
\cite{lim2020federated} & FL in Edge Networks & \checkmark & \checkmark & \text{\sffamily x} & \text{\sffamily x} & \text{\sffamily x} & \text{\sffamily x} \\ \hline
\cite{khan2021federated} & FL for IoT & \checkmark & \checkmark & \text{\sffamily x} & \checkmark & \text{\sffamily x} & \text{\sffamily x} \\ \hline
\cite{pham2021fusion} & FL for IIoT & \checkmark & \checkmark & \text{\sffamily x} & \text{\sffamily x} & \text{\sffamily x} & \text{\sffamily x} \\ \hline
\cite{xu2021federated} & FL for Health Informatics & \checkmark & \checkmark & \text{\sffamily x} & \checkmark & \text{\sffamily x} & \text{\sffamily x} \\ \hline
\cite{chen2022decentralized} & decentralized Wireless FL & \text{\sffamily x} & \checkmark & \text{\sffamily x} & \checkmark & \text{\sffamily x} & \checkmark \\ \hline
\cite{che2022decentralized} & DFL framework & \checkmark & \checkmark & \text{\sffamily x} & \text{\sffamily x} & \text{\sffamily x} & \checkmark \\ \hline
\cite{islam2022fbi} & Blockchain-based FL & \checkmark & \checkmark & \text{\sffamily x} & \text{\sffamily x} & \text{\sffamily x} & \checkmark \\ \hline
This work & DFL on the Edge & \checkmark & \checkmark & \checkmark & \checkmark & \checkmark & \checkmark \\ \hline
\end{tabular}
\end{table*}

\section{Wireless Mesh Network Performance Analysis}

In this section, an analytical approach is employed to assess the performance of a wireless mesh network, utilizing the physical interference model \cite{physical} to quantify the likelihood of successful data transmission between a transmitting node and a receiving node within the network. Our theoretical analysis draws heavily from principles of stochastic geometry and random point processes \cite{net, primer}. According to the physical interference model, the probability of successful transmission hinges on the Signal-to-Interference-and-Noise Ratio (SINR) observed at the receiver. A transmission is classified as successful if the SINR meets or exceeds a predefined threshold value. A notable advantage of the physical interference model is its comprehensive evaluation of total interference emanating from all nodes except the transmitter \cite{achievable_transmission}. Subsequently, we establish the network topology employing the Poisson point process (PPP) theory.

\subsubsection{Poisson Point Process}
In our analysis, we assume that the edge devices within the network are distributed based on a stationary homogeneous Poisson Point Process (PPP) with an intensity of $\lambda$. These devices are distributed within a disk $\mathcal{D}\subset\mathcal{R}^2$ with a radius of $R$, centred at the origin of the two-dimensional plane $\mathcal{R}^2$. According to the properties of a Poisson point process, the expected number ${N}$ of devices falling within the disk $\mathcal{D}$ can be calculated as ${N = \lambda|\mathcal{D}|}$, where $|\mathcal{D}| = \pi {R}^2$ \cite{net}. 

An important characteristic of the homogeneous PPP, as per Palm probability theory, is that adding an extra point at the origin in a specific realization of the process does not affect the distribution of the remaining points in the process (Slivnyak’s theorem). Consequently, interference statistics can be measured equivalently by assuming that the typical receiver is a point within the process located at the origin \cite{book}. In the network, each device is denoted by $i$ with $1 \leq i \leq N$, where $N$ represents the total number of active devices within the desired receiver coverage area. Additionally, ${i\in\varphi}$, which $\varphi$ encompasses all participants within the entire target area.

\subsubsection{Successful Transmission Probability}
In the subsequent analysis of the probability of successful transmission, a slotted ALOHA medium access control scheme is considered. In this scheme, each device independently decides to transmit with a probability of $p$, without coordination with other devices. Additionally, we assume Rayleigh fading for the propagation channel, where the transmission power of each device has a zero mean. The Signal-to-Interference-and-Noise Ratio (SINR) measured at the receiver located at the origin is calculated using the following equation:

\begin{equation}
    SINR = \frac{S}{\mathcal{N} + I}
\end{equation}

Here, $S$ represents the signal power emitted by the intended transmitter, $\mathcal{N}$ stands for the noise power, and $I$ corresponds to the cumulative interference power stemming from other transmitters.

For simplicity, we can theoretically assume that the noise power $\mathcal{N}$ is significantly lower than the total interference power. Therefore, we will employ the Signal-to-Interference Ratio (SIR) for the remainder of our analysis, as defined below:

\begin{equation}
    SIR = \frac{S}{I}
\end{equation}

The received signal power $S_i$ at the receiver from a transmitter $i$ is \cite{achievable_transmission, net}:

\begin{equation}
    S_i = P_ihr_{i}^{-\alpha}
\end{equation}

In this context, $P_i$ represents the transmission power of transmitter $i$, $h$ denotes the fading factor in accordance with the Rayleigh fading model, $r_i$ stands for the distance from transmitter $i$ to the origin, and $\alpha$ characterizes the path loss parameter, which reflects the attenuation of signal power with distance. The total interference power $I$ observed at the receiver results from the summation of all $S_i$ values, where $i$ corresponds to all transmitting devices except the intended transmitter. According to the Rayleigh fading model, the received signal power $S_i$ follows an exponential distribution \cite{rayleigh}, and with the assumption of unit transmission power, its distribution is defined by equation (\ref{eq: exp _4}) \cite{net}:

\begin{equation}
\label{eq: exp _4}
    f_{S_i}(x) = r_{i}^{\alpha}\text{exp}\bigg(-r_{i}^{\alpha}x\bigg),     \qquad    x\geq0
\end{equation}

The successful transmission probability, as per the physical interference model, can be described as the likelihood that the Signal-to-Interference Ratio (SIR) exceeds or equals a predefined threshold:
\begin{equation*}
     \begin{split}
    p_{succ} & = P(SIR\geq \theta) \\
              & = P(\frac{S}{I} \geq \theta) \\
              & = P(S \geq \theta I) \\
    \end{split}
\end{equation*}
\begin{equation}
\label{eq: exp _5}
     \begin{split}
    ~~~~~~~~~= \text{exp}(-\theta r^{\alpha} I)
    \end{split}
\end{equation}
where r is the distance between the desired transmitter and the receiver.

\subsubsection{Laplace Transform of Interference}

The successful transmission probability, as expressed in equation (\ref{eq: exp _5}), is equivalent to the Laplace transform of the cumulative interference observed at the receiver when evaluated at $(s = \theta r^{\alpha})$ \cite{net}.

\begin{equation}
\label{eq: lab _6}
    p_{succ} = \mathcal{L}_I(s) |_{s = \theta r^{\alpha}} = \mathbb{E}\big(\text{exp}(-sI)\big)|_{s = \theta r^{\alpha}}
\end{equation}

Following established analytical methods from stochastic geometry and probability-generating functional, as outlined in references \cite{net, primer}, we can deduce a closed-form expression for the successful transmission probability as follows:

\begin{equation*}
\begin{split}
    \mathcal{L}_I(s) & = \mathbb{E}\bigg(\text{exp}(-sI)\bigg) \\ 
    & = \mathbb{E}\bigg(\text{exp}(-s\sum_{x\in \Phi}hr_x^{-\alpha})\bigg)  \\
    & = \mathbb{E}_{\Phi}\left(\prod_{x\in \Phi} \mathbb{E}_{h}[ \text{exp}(-shr_x^{-\alpha})]\right) \\
\end{split}
\end{equation*}

This leads to the following expression for the successful transmission probability:

\begin{equation}
\label{eq: eq7}
    p_{succ} = \text{exp}\left(-\lambda p \pi R^2 s^{\frac{2}{a}} \frac{\frac{2\pi}{\alpha}}{sin(\frac{2\pi}{\alpha})}\right)
\end{equation}

Here, $p$ represents the probability of an individual transmitter deciding to transmit independently (ALOHA), ($\lambda$) stands for the intensity of the Poisson Point Process (PPP), $R$ signifies the radius of the PPP disk, ($r$) denotes the distance between the transmitter and the receiver, and ($\alpha$) represents the path loss parameter.

Consequently, from equations (\ref{eq: lab _6}) and (\ref{eq: eq7}), we can derive the following expression for the successful transmission probability:

\begin{equation}
    p_{succ} = \text{exp}\bigg(-\lambda p \pi R^2 r^2 \theta^{\frac{2}{\alpha}} \frac{\frac{2\pi}{\alpha}}{sin(\frac{2\pi}{\alpha})}\bigg) \label{eq: succ prob}
\end{equation}

\section{Methodology}
\label{sec: methodology}

Our system model encompasses various critical perspectives to optimize decentralized learning: We start from a system perspective, optimizing a decentralized model by collaborating between several distributed edge devices without direct access to their local data. Communication among devices occurs in a peer-to-peer manner, eliminating the need for a central server. From a spatial perspective, we leverage geometric patterns to efficiently manage multi-user communication. Each communication round identifies successful transmitter devices based on interactions with neighbours. Considering convergence, our approach incorporates theoretical analysis to define the target convergence state of the model. Furthermore, we introduce novel aggregator methods and employ Hidden Markov Models (HMM) for device evaluation. Historical performance guides the selection and weighting of edge devices in the learning process.
\subsection{System architecture}
In a typical Federated Learning (FL) training process, the following three key steps are involved. It's important to distinguish between the \textit{local model}, which is the model trained on each participating device, and the \textit{global model}, which is the model aggregated by the FL server. The following are the main learning steps:

\begin{enumerate}
    \item \textbf{Task Initialization (Step 1):} At the beginning, the server determines the training task, which refers to the specific application and its data requirements. The server also sets certain hyperparameters for the global model, such as the learning rate. Afterwards, the server shares the initial global model $\textbf{W}^0_G$ and the task details with selected participants.
    \item \textbf{Local Model Training and Update (Step 2):} Building on the current global model denoted as $\textbf{W}_t$ (with $t$ as the iteration index), each participant individually utilizes their local data and device to update their local model parameters, denoted as $\textbf{W}_t$. In each iteration, participant $i$ strives to find optimal parameters,  $\textbf{W}_t$, that minimize the loss function $L(.)$. The loss function varies depending on the problem and the model employed, as illustrated in the table (\ref{table:loss-functions}). In essence, they aim to find ${\textbf{W}_t^i}^*$ such that it minimizes $L(.)$. After updating, the local model parameters are then transmitted back to the server.
    \item \textbf{Global Model Aggregation and Update (Step 3):} In this final step, the server aggregates the local models received from all participants. After aggregating, the server generates updated global model parameters ${\textbf{W}^{t+1}_G}$ and sends them back to the respective data owners.
\end{enumerate}

In contrast to traditional Centralized Federated Learning (CFL) systems, the proposed Decentralized Federated Learning (DFL) architecture distinguishes itself by eliminating the need for a central aggregating server. Our target model revolves around a network of edge devices communicating through a wireless mesh infrastructure. The primary objective of this system is the collaborative optimization of parameters $\mathbf{W}$ for a global model represented as $\hat{y} = f(\mathbf{W}, x)$, where $\hat{y}$ represents the model's predicted output, and $x$ denotes input data. Each individual device possesses its distinct dataset $\mathcal{D}_i$ consisting of input data $x_i$, and this dataset remains private, not shared with other devices within the network.

The local loss function at each device can be defined as:
\begin{equation}
L_i(\mathbf{W}) = \frac{1}{|\mathcal{D}i|}\sum{x,y \in \mathcal{D}_i} l\bigg(\hat{y} - y; \mathbf{W}, x\bigg)
\end{equation}

In this equation, $|\mathcal{D}_i|$ represents the size of the local dataset, and $l(\hat{y} - y)$ is the loss function that quantifies the disparity between the model's predicted output and the actual output corresponding to input $x$. It is important to note that we assume the local loss function to be both convex and smooth.

At the outset of each training iteration $t$, let $\mathbf{W_{t_i}}$ represent the local model weights for each device $i$. Employing its local dataset, each device engages in Stochastic Gradient Descent (SGD) \cite{mcmahan2017communicationefficient} on the local loss function. The device subsequently updates its local model weights using the following equation:
\begin{equation}
\mathbf{W_{{t}_{i}+1}} = \mathbf{W{t}_i} - \mu \nabla L_i(\mathbf{W_{t_i}})
\end{equation}

Here, $\mu$ denotes the learning rate, carefully selected to ensure the convergence of the SGD algorithm to a minimum.

\begin{table}[ht]
\centering
\begin{tabular}{|c|p{0.6\linewidth}|}
\hline
\textbf{Model Name} & \textbf{~~~~~~~~~~~~~~~~~~Loss Function} \\
\hline
Neural Network \cite{spatial} & (classification) \[
L_i(\mathbf{W}) = \frac{1}{|\mathcal{D}_i|}\sum_{x,y \in \mathcal{D}_i} l\bigg(\hat{y} - y; \mathbf{W}, x\bigg) 
\]
\\
\hline
Linear Regression \cite{montgomery2021introduction} & 
Mean Squared Error (MSE): (Regression)
\[
L_{\text{LR}}(y, \hat{y}) = \frac{1}{2}(y - \hat{y})^2
\]

\\
\hline
K-means \cite{hartigan1979algorithm} & 
Sum of Squared Distances: (Clustering)
\[
L_{\text{K-means}}(x, c) = \sum_{i=1}^{K} ||x - c_i||^2
\]

\\
\hline
Squared-SVM \cite{suykens1999least}& 
Squared Hinge Loss: (Binary classification)
\[
L_{\text{SVM}}(y, \hat{y}) = \max(0, 1 - y \cdot \hat{y})^2
\]
\\
\hline
\end{tabular}
\caption{Loss functions for different models}
\label{table:loss-functions}
\end{table}

In the subsequent phase, each device transmits its recently updated local model weights $\mathbf{W_{{t}_{i}+1}}$ to its immediate one-hop neighbours within the wireless mesh network. Simultaneously, it receives updated local model weights from its corresponding one-hop neighbours. Subsequently, each device executes a local aggregation process on these received local model weights, typically involving straightforward averaging. This results in the creation of the initial updated local model weights that will be utilized in the subsequent iteration \cite{Savazzi_2020}.

The process outlined above facilitates the "diffusion" of each device's local model weight parameters throughout the network during each iteration. Essentially, this diffusion mechanism involves the dissemination of the impact of each device's local training dataset across the network through the transmission of local model weights. Notably, prior research, such as \cite{diffusion}, has demonstrated that this collaborative learning network converges to the same global optimum and at a similar convergence rate when compared to a conventional centralized cloud-server approach.

The main objective of DFL is to discover model parameters ($\textbf{W}$) that minimize the average loss function (also known as an object function or cost function) across all neighbour participating devices as follows:

\begin{equation}
\min_{\textbf{W} \in \mathbb{R}^d} [L(\textbf{W}_t) = \frac{1}{N} \sum_{i=1}^{N} \zeta_i L_i({\textbf{{W}_t}})] \label{eq: opt loss}
\end{equation}

In this context of the loss function in equation (\ref{eq: opt loss}), each device indexed as $i$ is assigned a weight denoted by $\zeta_i > 0$. In practical scenarios, these weights $\zeta_i$ are typically determined in proportion to the amount of data residing on each respective device. This means that devices with more data contribute more significantly to the overall objective, as represented by the optimization problem expressed in the equation (\ref{eq: opt loss}). Furthermore, we assume that the edge devices are capable of running the model within a certain time slot at each epoch.

\subsection{Communication system}

In our model, the Wireless Mesh Networks (WMNs) approach is proposed as a fundamental communication technique that allows devices to share their model parameters in a peer-to-peer manner. Our proposal results from the popularity that WMN have gained due to their cost-effectiveness, which makes them an attractive option for wireless connectivity in the DFL network. WMNs exhibit dynamic self-organization and self-configuration, allowing network nodes to establish and maintain mesh connections autonomously. This characteristic bestows several advantages upon WMNs, including low initial expenses, efficient network upkeep, resilience, and consistent service coverage \cite{akyildiz2005wireless}. In addition, this low-cost WMN infrastructure is well-suited for establishing a DFL network that can span across community networks, metropolitan areas, municipalities, and enterprise networks.

\subsection{Learning convergence criterion}

Consider a scenario with $N$ edge devices participating in the learning network. For this analysis, we make the assumption that these edge devices are distributed according to a homogeneous PPP (Poisson Point Process) with an intensity measure denoted as $\lambda$. Furthermore, these devices are confined within a circular region $\mathcal{D}$ centred at the origin and having a radius of R. Within the scope of a typical receiver positioned at the origin, the probability of successful data transmission from a transmitter located at a distance $r$ from the origin can be determined using Equation (\ref{eq: succ prob}).

The devices that successfully transmit data to the receiver also follow a homogeneous PPP, but with an intensity measure of $\lambda p_{\text{succ}}$ \cite{spatial}. Consequently, the number of devices that succeed in transmitting their data to the receiver, denoted as $\tilde{N}$, can be expressed as:

\begin{equation}
\tilde{N} = |\mathcal{D}|\lambda p_{\text{succ}} = \pi R^2 \lambda \exp\left(-\lambda p \pi R^2 r^2 \theta^{\frac{2}{\alpha}} \frac{\frac{2\pi}{\alpha}}{\sin\left(\frac{2\pi}{\alpha}\right)}\right)
\end{equation}

The distribution of the number $m$ of devices successfully transmitting can be described as \cite{spatial}:

\begin{equation}
P(m = l) = \frac{\exp(-\tilde{N}) \tilde{N}^l}{l!}
\end{equation}

Let $m_{t,j}$ represent the number of devices successfully transmitting their updated local model weights to the receiving device $j$ during training iteration $t$, resulting in the global model $\textbf{W}_j$. Additionally, let $N_j$ denote the total number of training iterations out of $t$, in which at least one device successfully transmitted local model weights to receiver $j$. We can now introduce the convergence condition for the model training procedure within the DFL network, adapted from \cite{spatial} and configured to suit the described decentralized architecture:

\begin{equation}
    \underset{\textbf{W}_j}{\text{arg\,min}}~\bigg(\frac{1}{N_j} \sum_{j=1}^{N_j} \|\nabla L_j(\mathbf{W_t})  \|\bigg) \leq \epsilon_0
\end{equation}

The learning convergence condition can be expressed as follows: The network achieves convergence after $R$ training iterations when the maximum expectation of the average gradient, taken across all participating devices denoted as $N_j$, does not exceed a predefined convergence threshold $\epsilon_0$. This expectation is calculated with respect to the distribution of the input dataset. In essence, this convergence criterion ensures that if even the device with the poorest performance, in terms of the expected average gradient after $R$ rounds, meets the convergence threshold, then all other devices should also meet it. In such a scenario, the DFL network is considered to have converged to the optimal model weight parameters.

\subsection{Device selection and models aggregator method}

In the context of FL, participant selection is a crucial aspect as it determines which edge clients or devices in the network will contribute to the collaborative model training process. In the literature, different methodologies are used to evaluate the distributed devices and choose the most appropriate group based on the required purpose. In FL, Hidden Markov Models (HMMs) \cite{eddy1996hidden}, which are probabilistic models widely used in various fields,  can be utilized to make well-informed choices concerning participant selection by modelling the past behaviour and performance of devices and thus regularly assign the weights ($\zeta_i$) in the equation (\ref{eq: opt loss}) for each connected device.

In the DFL approach, the master devices responsible for aggregating models from other neighbours at iteration $t$ must meet specific specifications and requirements to efficiently manage the learning process during that iteration.

Initially, the central server in CFL or master device in the DFL approach initializes a global model \(w_0\) randomly. Subsequently, in each communication round, the following sequence of steps is executed to achieve the learning objective, as illustrated in Figure (\ref{fig2}):
\begin{enumerate}
    \item Step I: Broadcast Latest Model. The central server (in CFL scenario) or the master device (in DFL scenario) disseminates the most recent global model \(w_t\) to all clients (neighbours)  (typically in cross-silo FL) or a subset of clients ($N_t$) chosen for participation in the current training round (commonly in cross-device FL).
    \item Step II: Clients Compute Local Updates. Each client (i.e., edge device) utilizes its compressed proposal model to calculate the model update based on its local dataset by performing multiple iterations of gradient descent: \(w_{t+1}^i \leftarrow w_{t+1}^i - \eta \nabla_w L(w_{t}^i,D_i)\), with \(\eta\) representing the learning rate.
    \item Step III: Aggregate Client Updates. The server or the master device updates the global model by combining the local updates using a specific aggregation rule $A(\cdot)$: $(w_{t+1} \leftarrow A(\{w_{t+1}^i: i\in\varphi\})$.
\end{enumerate}

The most widely-used aggregation rule for communication-efficient Federated Learning (FL) is Federated Averaging (FedAvg) \cite{mcmahan2017communicationefficient}, which aggregates the client updates by computing a weighted average:
\begin{equation}
 w_{t} \leftarrow \frac{1}{N_t}\sum_{i\in\varphi} w_{t}^{i}  
\end{equation}

However, FedAvg is not fault-tolerant, and even a single faulty/malicious client can prevent the global model from converging \cite{yin2018byzantine}. Although this work does not specifically address malicious attacks and model protection, it is important to mention that there are existing robust aggregation techniques designed for these purposes:

\textbf{Krum}, as described in \cite{blanchard2017machine}, operates in each communication round by selecting $m$ local model updates out of the total available $N_t$ updates to compute the global model update. This selection is based on comparing the similarity between these local updates. Assuming we have $f$ clients out of $N_t$ are malicious, Krum assigns a score to each local model update $w_i$ by calculating the sum of Euclidean distances between $w_i$ and the $m$ nearest neighbouring local updates among $N_t - f - 2$. The $m$ local model updates with the smallest scores are then chosen, and their average is calculated to determine the global model update.

\textbf{Median} \cite{yin2018byzantine}: The Median aggregation method is a coordinate-wise aggregation rule that operates independently on each model parameter. To determine the $i$th parameter of the global model update, the server arranges the $i$th parameter values from the submitted $N_t$ local model updates in ascending order and selects the median value. Median aggregation can achieve an order-optimal statistical error rate when dealing with strongly convex loss functions.

Various other aggregation methods have been proposed apart from the previously mentioned ones. For example, Bulyan \cite{guerraoui2018hidden} employs an iterative approach with aggregation rules like Krum for enhanced robustness, but it suffers from computational inefficiency and lacks scalability. Zeno \cite{xie2019zeno} assigns scores to updates and aggregates the top $N_t - b$ updates with the highest scores, where $N_t$ represents the total number of clients, and $b$ is a predefined hyperparameter, typically set equal to or greater than the number of malicious clients. 

Another recent approach uses variational autoencoders to project client updates into a latent space for malicious update detection, but it relies on the unrealistic assumption of having access to data matching the client's private data distribution for training. Some studies focus on achieving robust federated learning by identifying and blocking malicious clients through adaptive model quality estimation \cite{munoz2019byzantine} or clustered federated learning \cite{sattler2020byzantine}.

However, it is important to note that some of these methods have predominantly been tested in the context of centralized federated learning, specifically in the straightforward cross-silo scenario. Their applicability to the more complex and dynamic cross-device scenarios, which are characteristic of DFL approaches, has seen limited exploration. This limitation is part of our work to investigate.

\section{Model Simulation}
\label{sec: Simulation and Experiments}

The Simulation section of this work encompasses the following components:

\begin{enumerate}[(i)]
    \item Simulation of a Wireless Mesh Network: We simulate a wireless mesh network in which participants are distributed according to a Poisson Point Process (PPP), as discussed in our theoretical analysis of wireless mesh network performance. This simulation allows us to model communication dynamics and evaluate network performance across various scenarios.

    \item Simulation of the baseline Centralized Federated Learning (CFL) architecture: Our simulations replicate the baseline centralized FL architecture. In this setup, a multi-layered Convolutional Neural Network (CNN) is subjected to compression using the state-of-the-art genetic algorithm-based approach (SOTA compression method). The EMNIST benchmarking dataset \cite{cohen2017emnist} is employed for training and validation. The primary objective is to minimize communication overhead while enhancing the performance of the compressed CNN model for categorical digit classification with the existence of a central server. This improvement is achieved by enabling collaborative learning among multiple participants within the CFL framework.

    \item The simulations evaluated the proposed DFL architecture, which incorporates a state-of-the-art compression method. Specifically, a multi-layered Convolutional Neural Network (CNN) was compressed using a genetic algorithm-based approach at the edge. Additionally, various aggregation methods, such as Krum and Median, were employed, mirroring the approach used in the centralized setup. These simulations facilitated a performance comparison between the centralized and decentralized architectures, allowing us to assess the effectiveness and potential advantages of the DFL approach, which has no need for a central server and network infrastructure, taking into account the communication overhead and the complexity. Importantly, this work introduces the novel application of Krum and Median aggregation methods within the realm of DFL.

\end{enumerate}

Through these simulations, we aim to gain insights into the performance, accuracy, and efficiency of both CFL and DFL architectures within the context of wireless mesh networks. The simulation results will provide valuable information for understanding the feasibility and practical implications of implementing DFL in real-world scenarios.

Moreover, the primary objective of the simulation component in this project was to explore the relative performance of two FL architectures, (a) conventional centralized (CFL) and (b) fully decentralized (DFL), while considering the success probability of each data transmission between any two network participants. Notably, this work represents the first instance of providing system simulations where each communication step is completed with a specific success probability, accounting for the underlying communication network's performance within the physical interference model. These results measure the realistic performance of practical FL systems, avoiding the assumption of faultless communications.

To incorporate the communication success probability parameter into the algorithms for both FL architectures, we encountered challenges. Existing FL software libraries, such as TensorFlow Federated \cite{abadi2016tensorflow}, and PySyft \cite{ryffel2018generic}, do not offer the flexibility to define precisely how that network participants communicate during the learning process. Therefore, we developed a fully customized software solution based on the FedML FL and torch frameworks, which provided the most flexibility in defining and implementing custom FL network designs \cite{he2020fedml}. Specifically, after acquiring average successful transmission probabilities for various wireless mesh networking configurations through corresponding simulations, we integrated this parameter into the FL systems (i.e., CFL and DFL systems) simulation. This approach effectively incorporates the communication network aspect into the learning procedure.

In the following subsections, we provide further detailed descriptions of the simulation setups in other related aspects and present the corresponding results.

\subsection{Wireless Mesh Network Simulation}

To simulate the wireless mesh network that constitutes the backbone for communications across both centralized and DFL architectures, we assume that the participating edge devices are placed randomly in a bounded two-dimensional area of a circle according to a PPP. The communication medium is characterised by an inverse square path loss law and by the presence of Rayleigh fading modelled by an exponential random variable.

In the case of the proposed decentralized architecture, it is assumed that each edge device transmits and receives local model updates only from one-hop neighbours with regard to the wireless mesh network connectivity graph. The simulation of this one-hop neighbourhood is implemented by considering a disk $\mathcal{D}$ of radius $r_{oh}$ around the typical receiver and including a communication link for every transmitting device that falls inside this area. In other words, each edge device transmits to other devices with a distance less than $r_{oh}$. Moreover, the medium access control (MAC) scheme is Slotted Aloha with the probability of transmission $p$ in each slot. It is further assumed that there are $k$ available frequency levels spanning the allocated bandwidth, which are used for concurrent transmissions from transmitters to receivers, with $k = \lambda p \pi r_{oh}^2$ equal to the mean value of the number of transmitting devices falling inside the one-hop neighbourhood disk area of the typical receiver. Therefore, the desired one-hop-neighbouring transmitting devices do not interfere with each other with respect to the typical receiver.

As previously stated, the aim of the network simulation is to find the mean value of the successful transmission probability between a transmitting node placed inside the area of the circle and the typical receiver situated without loss of generality at the origin of the two-dimensional plane. In order to calculate the target probability, the following procedure is followed: 

\begin{enumerate}
    \item Generate a Poisson distributed for a random number of edge devices inside a disk of radius $R$ with a mean value equal to the process intensity $\lambda$ times the area of the disk $A = \pi R^2$.
    
    \item Split all generated edge devices into transmitters and receivers for a specific round of communications, i.e. perform thinning of the main PPP with parameter $p$.
    
    \item For a typical receiver placed at the origin, find the number of transmitters that are less than $r_{oh}$ distance away.
    
    \item Calculate the Signal to Noise and Interference Ratio (SINR) for each of the transmitting devices, considering as interference all transmitters that are outside the one-hop neighbourhood radius. If the SINR is above the threshold $\gamma$ the transmission is considered successful.
    
    \item Repeat the above steps for $N$ rounds and calculate the successful transmission probability for a specific threshold $\gamma$ as the total number of successful transmissions over the total number of transmissions.
\end{enumerate}

\subsection{centralized Federated Learning (CFL)}
In simulating the fundamental CFL architecture as proposed in \cite{mcmahan2017communicationefficient}, the central server node is assumed to be positioned at the origin of the two-dimensional plane, as previously described in our wireless mesh network simulation setup. Within this setup, the edge devices participating in the learning process are located within a one-hop neighbourhood region of a disk, utilizing some distinct frequency bands for simultaneous transmission. Each training iteration involves the following sequence of actions:
\begin{itemize}
    \item The central server disseminates the global model to all participating devices.
    \item Every edge device employs Stochastic Gradient Descent with its own local dataset to adjust the weights of the convolutional neural network and subsequently uploads these updated weights to the central server.
    \item Finally, the server consolidates the received local models from each edge device using the Federated Averaging, Krum and Median aggregator aggregators algorithm, applying straightforward one of these aggregators methods for the global model update.
\end{itemize}

Following is a description of the convolutional neural network used for classifying handwritten digits trained on the EMNIST dataset. The input of the neural network is a 28$\times{28}$ pixels image depicting a handwritten digit in grayscale. The model is compressed using the output layer consisting of 10 outputs corresponding to each handwritten digit, whereas the genetic algorithm is used to minimize the overall model size. The EMNIST benchmarking dataset \cite{cohen2017emnist} contains 60K grayscale images of handwritten digits. For simulation purposes, the dataset is split into random-sized parts and allocated among the participating edge devices.

\subsection{Decentralized Federated Learning (DFL)}

In order to facilitate a meaningful comparison of results, the simulation setup for the fully decentralized FL system mirrors the configuration of the baseline centralized system, maintaining consistent setup parameters such as learning rate, stochastic gradient descent batch size, and the density of participating devices. Analytically, for a given edge device density (the total number of participating devices in the learning network), the objective is to train the convolutional neural network with the EMNIST benchmark dataset evenly distributed among all participating devices. Below, we outline the typical training iteration process for the fully decentralized algorithm:

\begin{itemize}
    \item Initialization: At the onset of each training round, each local edge device possesses the current convolutional neural network weights.

    \item Local Model Update: Each device independently performs stochastic gradient descent to update its local model weights using its local dataset.

    \item Communication and Model Exchange: Following the identification of one-hop neighbours within the wireless mesh network connectivity graph for the specific training iteration (as previously noted, this graph is dynamically reconfigurable), each device initiates a broadcast. During this broadcast, it shares its intermediate updated local model with these neighbouring devices and, in turn, receives updated local models from them.

    \item Aggregation with Robust Methods: Each edge device aggregates the received local models with its own, employing a straightforward non-weighted average of each model weight, akin to Federated Averaging for the centralized case. Additionally, our work extends beyond traditional aggregation methods and evaluates the model using various robust aggregator methods, such as Krum and Median. This multifaceted approach aims to enhance performance and resilience in the face of adversarial behaviour, contributing to the robustness of the DFL framework.
\end{itemize}

This training iteration cycle, orchestrated collaboratively by the participating edge devices, enables the fully DFL system to continually refine its global model. Importantly, it mirrors the core concept of FL, leveraging the collective knowledge of decentralized devices while preserving data privacy and security. The synchronization and aggregation of local models among neighbouring devices foster collaborative learning without the need for a centralized server, emphasizing the robustness and decentralized nature of this approach.

\begin{figure}[t]
\begin{centering}
\includegraphics[width=3.3in,height=2.2in]{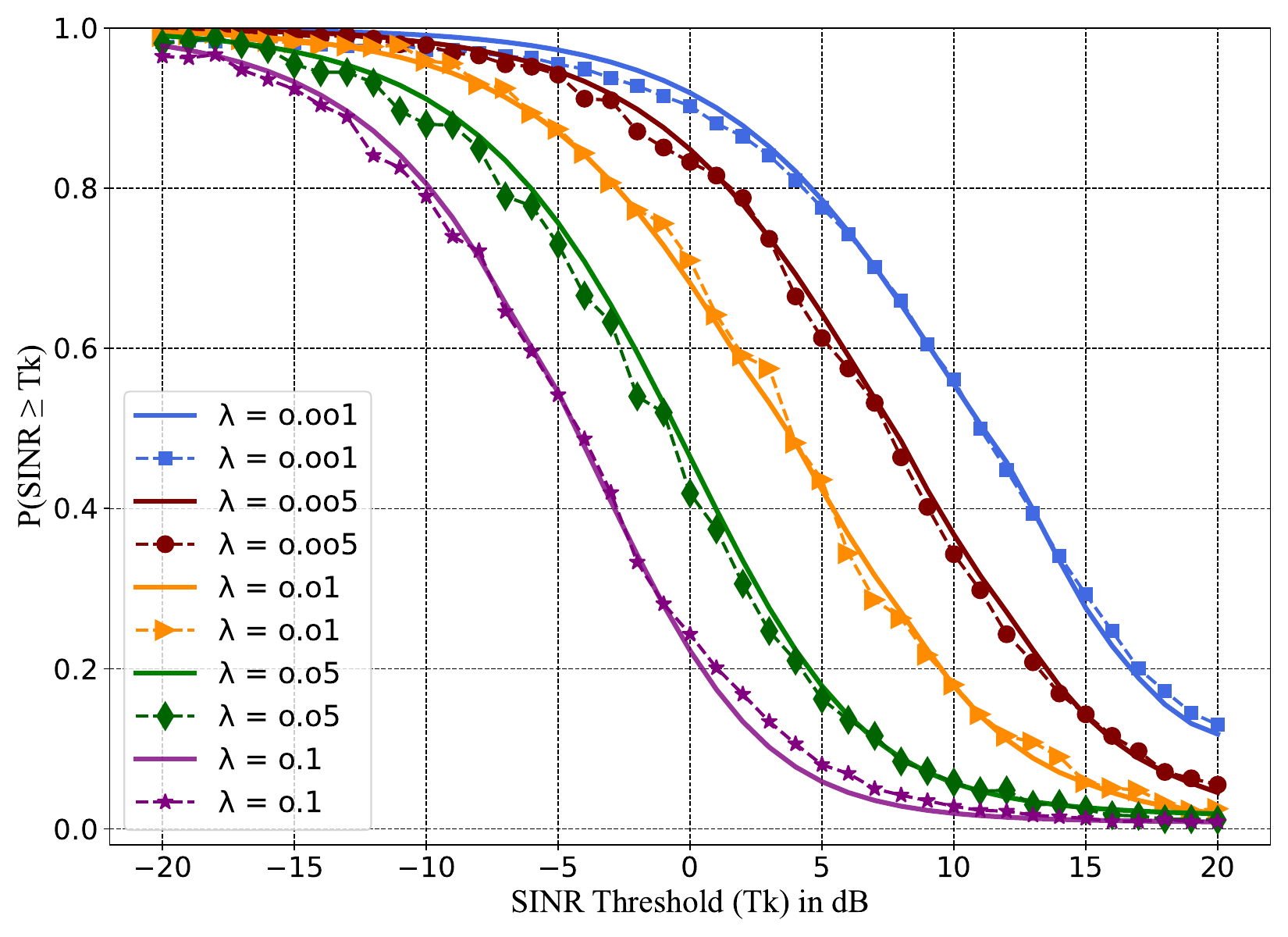}
\par\end{centering}
\caption{The theoretical (solid lines) and simulation (dashed with markers) successful transmission probability as a function threshold values $\gamma$ in dB.}
\label{fig4}
\end{figure}

Furthermore, the powerful benefit in our DFL model, particularly in high-intensity networks, is designed to leverage the collective computational power of edge devices efficiently. By distributing tasks, minimizing data transfer, and promoting parallel processing, this approach inherently leads to reduced latency, making it well-suited for scenarios where low-latency responses are essential.

\section{Results and Discussion}
\label{sec: results and discussion}

In this section, we provide a detailed overview of the simulation parameters and present the results of our wireless mesh network simulation.

For the wireless mesh network simulation, we selected a disk radius of $R = 500$ meters, defining the total simulation area. The intensity of the Poisson Point Process (PPP), representing the density of participating edge devices, was varied, with values $\lambda = 10^{-3}, 5\times10^{-3}, 10^{-2}, 5\times10^{-2}, 10^{-1}$, resulting in average numbers of participating devices ($\tilde{N}$) is varied. 

To emulate the behaviour of devices following the slotted Aloha Medium Access Control (MAC) scheme, we set the probability of a transmitting device deciding to transmit in a specific slot to $p = 0.3$. The one-hop neighbourhood disk radius was set to $r_{oh} = 200$ meters.

We conducted SINR calculations for several threshold ($\gamma$) values, spanning from -20 dBm to +20 dBm, with $N_{sim} = 10^5$ simulation rounds for each threshold setting. Additionally, we assumed that each device transmitted with the same power level of $P_{tx} = 1$ Watt.

The diverse set of simulation parameters employed here enables us to comprehensively explore the performance of our wireless mesh network under various conditions. By systematically varying $\lambda$ and $\gamma$, we gain insights into the network's robustness, scalability, and reliability, shedding light on its behaviour across different device densities and signal-to-noise environments.

This detailed analysis serves as a strong foundation for our subsequent discussion and allows us to draw meaningful conclusions about the suitability of our proposed approach for real-world scenarios.

With regards to the FL network training parameters for both centralized and decentralized architectures, the EMNIST dataset was initially split into training and validation sets, comprising 50,000 and 10,000 samples, respectively. In configuring the stochastic gradient descent algorithm's hyperparameters, we selected a learning rate ($\mu$) of 0.015 and a batch size ($n_{batch}$) of 32 samples. The performance metrics under consideration include model accuracy and cross-entropy loss, specifically tailored for categorical data.

In Figure \ref{fig4}, we present the calculated successful transmission probability as a function of varying $\gamma$ threshold values, encompassing different PPP intensity values in both theoretical and simulation contexts. A discernible trend emerges, showcasing that as the participating device density increases, the resulting mean value of the successful transmission probability experiences a non-linear decrease. This relationship can be attributed to the escalating total interference power at the typical receiver, resulting from more devices transmitting data, driven by a constant slotted Aloha transmission probability ($p$).

\begin{figure}[t]
\begin{centering}
\includegraphics[width=3in,height=2 in]{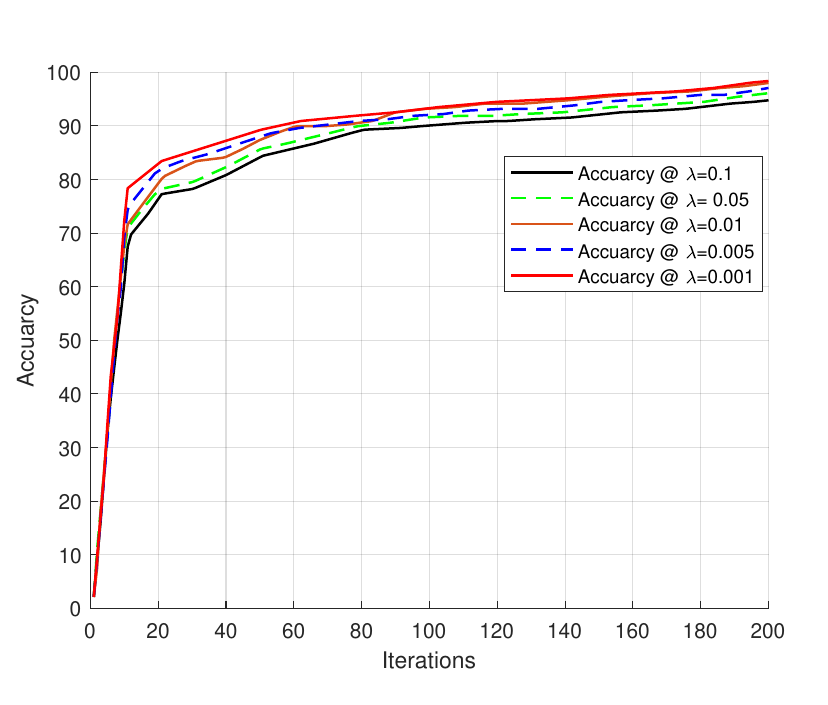}
\par\end{centering}
\caption{Model accuracy as a function of training iterations for CFL.}
\label{fig5}
\end{figure}
\begin{figure}[t]
\begin{centering}
\includegraphics[width=3.in,height=2in]{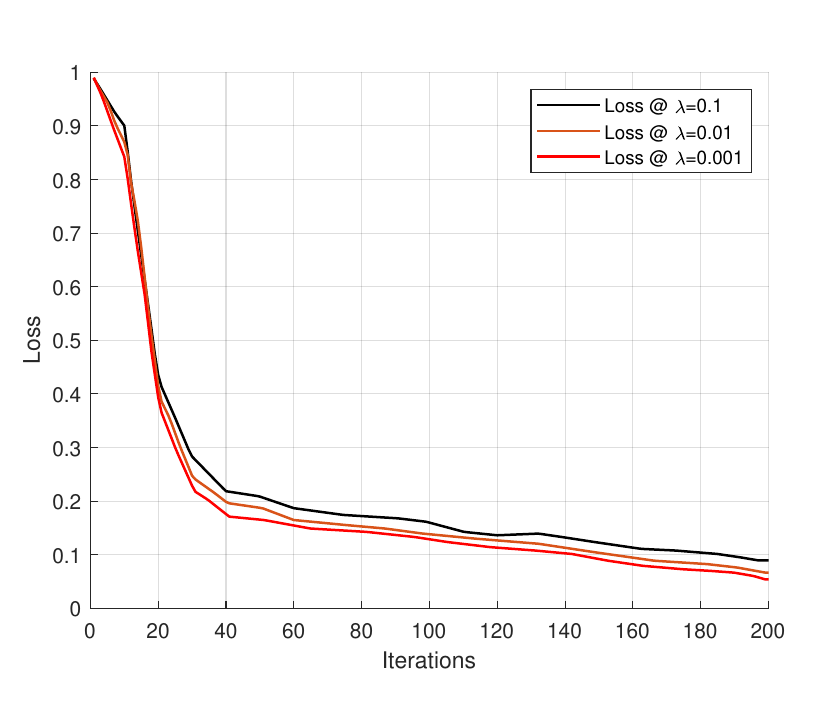}
\caption{Model cross-entropy loss as a function of training iterations for CFL.}
\label{fig6}
\end{centering}
\end{figure}
\begin{figure}[t]
\begin{centering}
\includegraphics[width=3in,height=2in]{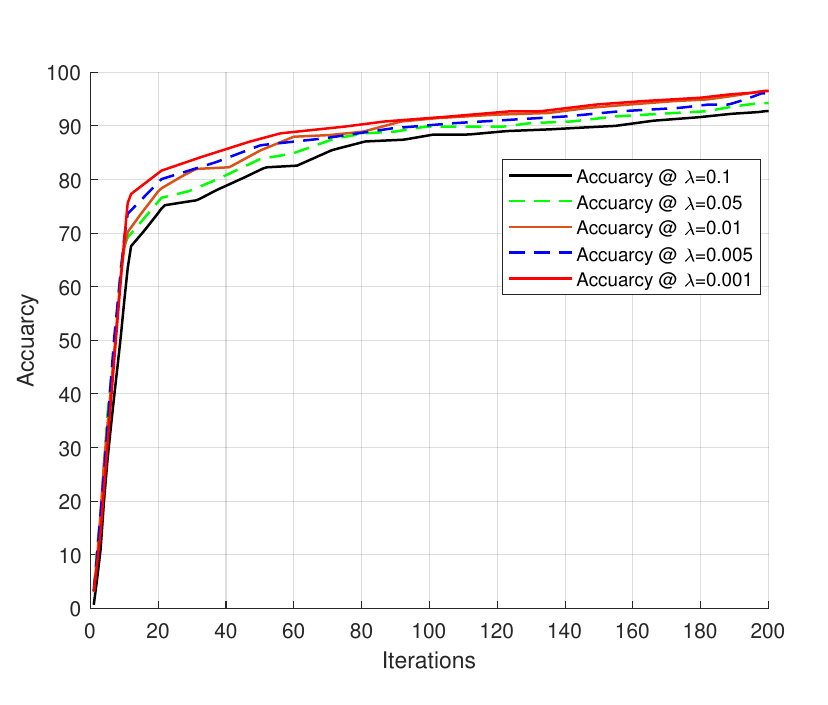}
\caption{Model accuracy as a function of training iterations for DFL.}
\label{fig7}
\end{centering}
\end{figure}
\begin{figure}[t]
\begin{centering}
\includegraphics[width=3 in,height=2in]{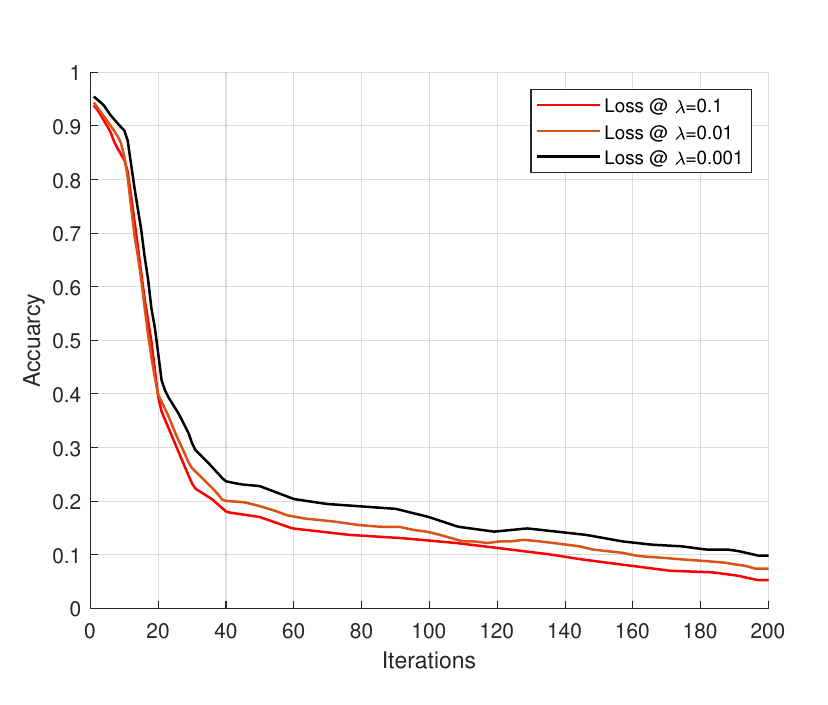}
\caption{Model cross-entropy loss as a function of training iterations for DFL.}
\label{fig8}
\end{centering}
\end{figure}

It's noteworthy that, in both CFL and DFL systems, we adopted an SINR threshold ($\gamma$) value of -16 dBm as the criterion for deeming a transmission successful. This choice of threshold ensures that our evaluation remains consistent across different scenarios, providing a clear basis for comparative analysis.

Below in table \ref{tab1}, the convolutional neural network model accuracy and cross-entropy loss for centralized and DFL architectures are presented for increasing model training epochs and device density $\lambda = 0.01$.

The simulations were conducted in a 20-fold manner, and the average values for each metric are presented here. In the case of decentralization, performance metrics are derived by averaging the individual metrics of each participating device across the entire network. It is worth noting that in the centralized scenario, the model achieves a convergence threshold of over 95\% model accuracy after an average of 152 training epochs. Conversely, in the decentralized scenario, the predefined threshold is reached after an average of 173 epochs. This represents a 13.8\% increase in the required number of training iterations, consequently impacting the overall system latency to the same degree.

\begin{table}[t]
\caption{FL Simulation Results}
\label{tab1}
\centering
\begin{tabular}{ccccc}
\toprule
\textbf{Iterations} & \multicolumn{2}{c}{\textbf{CFL}} & \multicolumn{2}{c}{\textbf{DFL}} \\
\cmidrule(lr){2-3} \cmidrule(lr){4-5}
& \textbf{Accuracy} & \textbf{Loss} & \textbf{Accuracy} & \textbf{Loss} \\
\midrule
10 & 0.7987 & 0.8743 & 0.7764 & 0.9321 \\
20 & 0.8453 & 0.4422 & 0.8351 & 0.5276 \\
30 & 0.8662 & 0.2917 & 0.8497 & 0.3344\\
40 & 0.8877 & 0.2587 & 0.8723 & 0.2897\\
50 & 0.9012 & 0.2492 & 0.8903 & 0.2632\\
75 & 0.9249 & 0.2314 & 0.9122 & 0.2471\\
100 & 0.9388 & 0.2265 & 0.9276 & 0.2384\\
125 & 0.9464 & 0.2075 & 0.9304 & 0.2172\\
150 & 0.9497 & 0.1794 & 0.9415 & 0.1964\\
175 & 0.9753 & 0.1623 & 0.9511 & 0.1742\\
200 & 0.9892 & 0.1428 & 0.9759 & 0.1515\\
\bottomrule
\end{tabular}
\end{table}

\begin{figure*}[t]
\centerline{\includegraphics[scale=0.7]{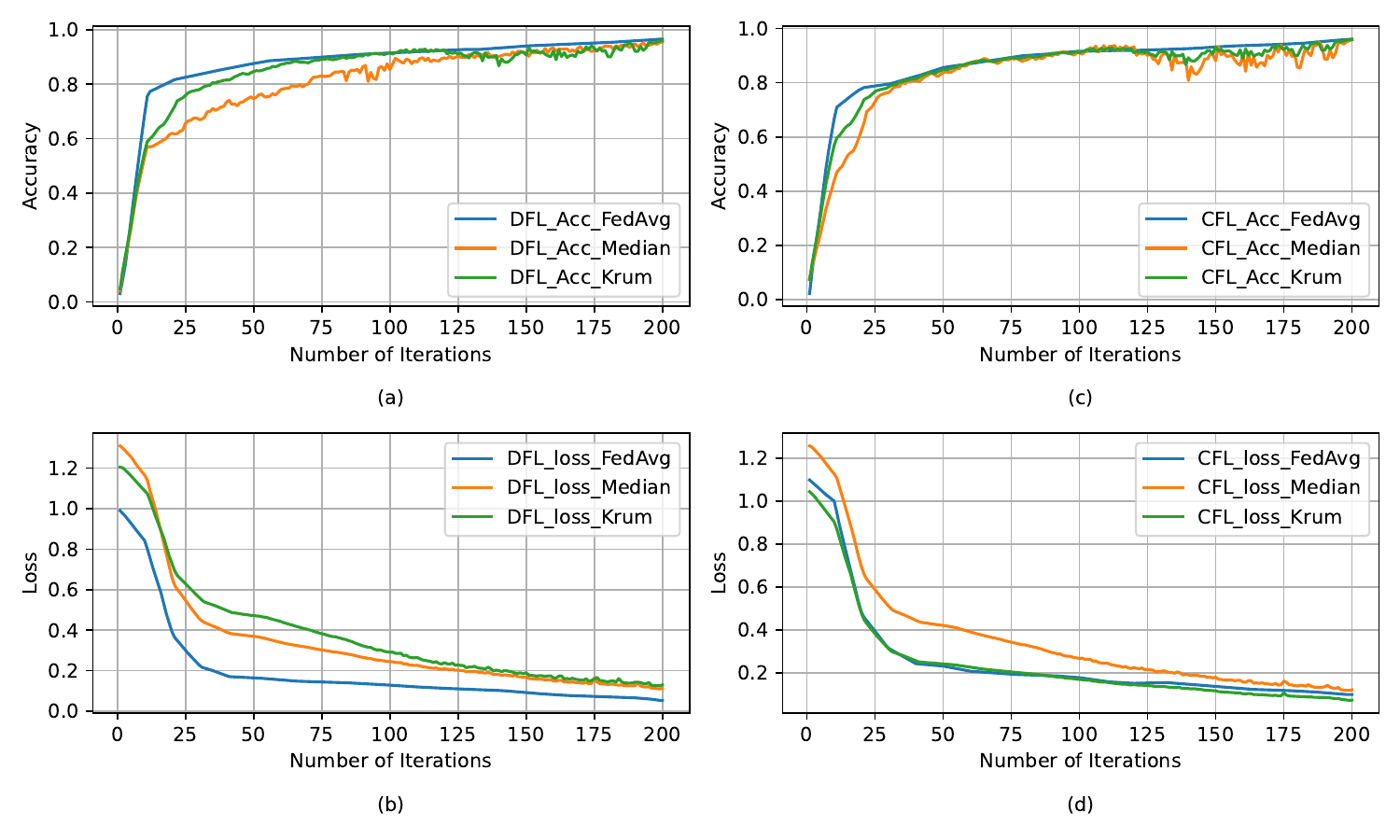}}
\caption{Comparative analysis of DFL and CFL performance metrics using various aggregator methods.}
\label{fig_DFL_CFL}
\end{figure*}

The increase in training epochs can be attributed to the diffusion delay introduced by the decentralized architecture. In the decentralized approach, the contributions of each device's local dataset to the local model update are not immediately transferred across the network. This is in contrast to the centralized approach, where a single aggregator collects all individual contributions in each round to update the global model. Thus, this trade-off becomes evident when transitioning to a fully decentralized solution with a realistic model of the core communication network.

Despite the trade-off of increased training epochs, the advantages of data security and resilience make DFL a compelling choice for collaborative machine learning in decentralized settings. This phenomenon highlights the practical modelling of the core communication between devices in DFL scenarios, where the network topology and communication mechanisms play an important role in shaping the learning dynamics. Understanding and addressing these challenges are essential steps toward optimizing DFL frameworks for real-world applications.

Figures \ref{fig5}, \ref{fig6}, \ref{fig7}, and \ref{fig8} illustrate the accuracy and cross-entropy loss of both architectures concerning varying device densities ($\lambda$). These charts clearly demonstrate an inverse relationship between model accuracy and network device density. This outcome emphasizes the significance of considering the communication network's performance in the analysis of an FL system. Interestingly, an increased number of devices participating in the learning process, while expected to enhance convergence, actually leads to reduced model accuracy due to elevated interference from additional transmitters.

Furthermore, it is noteworthy that the rate of increase in accuracy until the 90\% threshold is nearly identical between the baseline centralized solution and the proposed DFL system. This observation is of particular significance, indicating that by slightly adjusting the convergence threshold criterion, there's no compromise in performance when transitioning to the proposed DFL framework.

Our model employs a genetic algorithm-based approach to optimize model size, making it suitable for resource-constrained IoT and wearable devices. This model compression contributes to reducing the complexity as the model size decreases, enhancing its applicability in such constrained environments. Despite this reduction and local model compression to minimize communication overhead and complexity, our approach excels in achieving high model performance. It competes effectively with traditional DFL models, as demonstrated in Figure (\ref{fig_DFL_CFL}). This highlights the efficiency and promise of our approach in striking a balance between model size and performance.

Our study conducts a comprehensive evaluation of the proposed model, considering various aggregator methods applicable across a spectrum of use cases. Notably, we integrate Median and Krum aggregator methods into the DFL framework, making our work a pioneering effort in this regard. Figure \ref{fig_DFL_CFL} presents a comparative analysis of DFL and CFL performance metrics over multiple training iterations, employing a diverse set of aggregator methods, including FedAvg, Median, and Krum. Subplot (a) provides insight into DFL's accuracy trends, while subplot (c) showcases CFL's accuracy trends. Subplots (b) and (d) delve into the corresponding loss trends for DFL and CFL, respectively. This in-depth analysis offers a window into the intricacies of training dynamics and convergence behaviour inherent to both approaches.

Lastly, our results consistently demonstrate that the DFL model consistently achieves accuracy rates exceeding 93\% and exhibits lower loss across all our proposed aggregator methods. This makes it a versatile and suitable model for various purposes. These findings emphasize the inherent advantages of DFL, especially in decentralized settings where concerns about data privacy and distribution are paramount. Moreover, they underscore DFL's potential as a robust and versatile framework perfectly suited for collaborative applications across edge devices.

\section{Conclusion and Future Work}
\label{sec: Conclusion and Future Work}

In conclusion, our study explores a contemporary approach to distributed machine learning, providing an alternative to traditional Internet of Things (IoT) model learning systems. We prioritize data security by keeping raw edge device data local and sharing only model parameters. Recognizing limitations in CFL systems, such as single points of failure and susceptibility to malicious attacks, we embrace a DFL architecture. 

Our work addresses DFL framework limitations through convergence analysis and comprehensive performance assessment of the communication network, utilizing wireless mesh networking. Our theoretical analysis employs stochastic geometry to derive a closed-form equation for successful transmission probability in wireless mesh networks. We present a detailed DFL architecture description and training procedure, establishing a learning convergence criterion. Simulations compare our DFL architecture to the conventional CFL model. Using practical slotted Aloha wireless mesh networks and the EMNIST dataset for handwritten digit classification, we extract the successful transmission probability, accounting for potential transmission failures.

Results demonstrate that our decentralized architecture closely matches centralized counterparts in terms of accuracy and average loss. Importantly, our study integrates geometric analysis and diverse aggregator methods (Krum and Median) over compressed models, achieving high performance while significantly reducing the communication overhead. This approach highlights the practicality of decentralized architectures and offers an efficient framework for future IoT systems, potentially scalable in real-world applications.

Future work for this project will focus on the following areas:

\begin{enumerate}[(i)] 

    \item Investigation of suitable networking protocols: Further research will explore networking protocols specifically tailored for wireless mesh networks, such as the Thread protocol. The aim is to identify protocols that support low-power, low-data rate transmissions, which are particularly well-suited for IoT system solutions. This research will contribute to the development of more efficient and optimised communication mechanisms within wireless mesh networks.

    \item Exploration of blockchain-enabled solutions: Future research could investigate the integration of blockchain technology into the compressed DFL architecture. As blockchain technology matures, it presents opportunities for enhancing data security, privacy, and trust in the context of FL. The exploration of blockchain-enabled solutions can contribute to the development of more robust and resilient DFL systems, providing additional layers of data integrity and transparency.
\end{enumerate}


\bibliographystyle{IEEEtran}
\bibliography{output}

\begin{IEEEbiography}[{\includegraphics[width=1in,height=1.15in,clip]{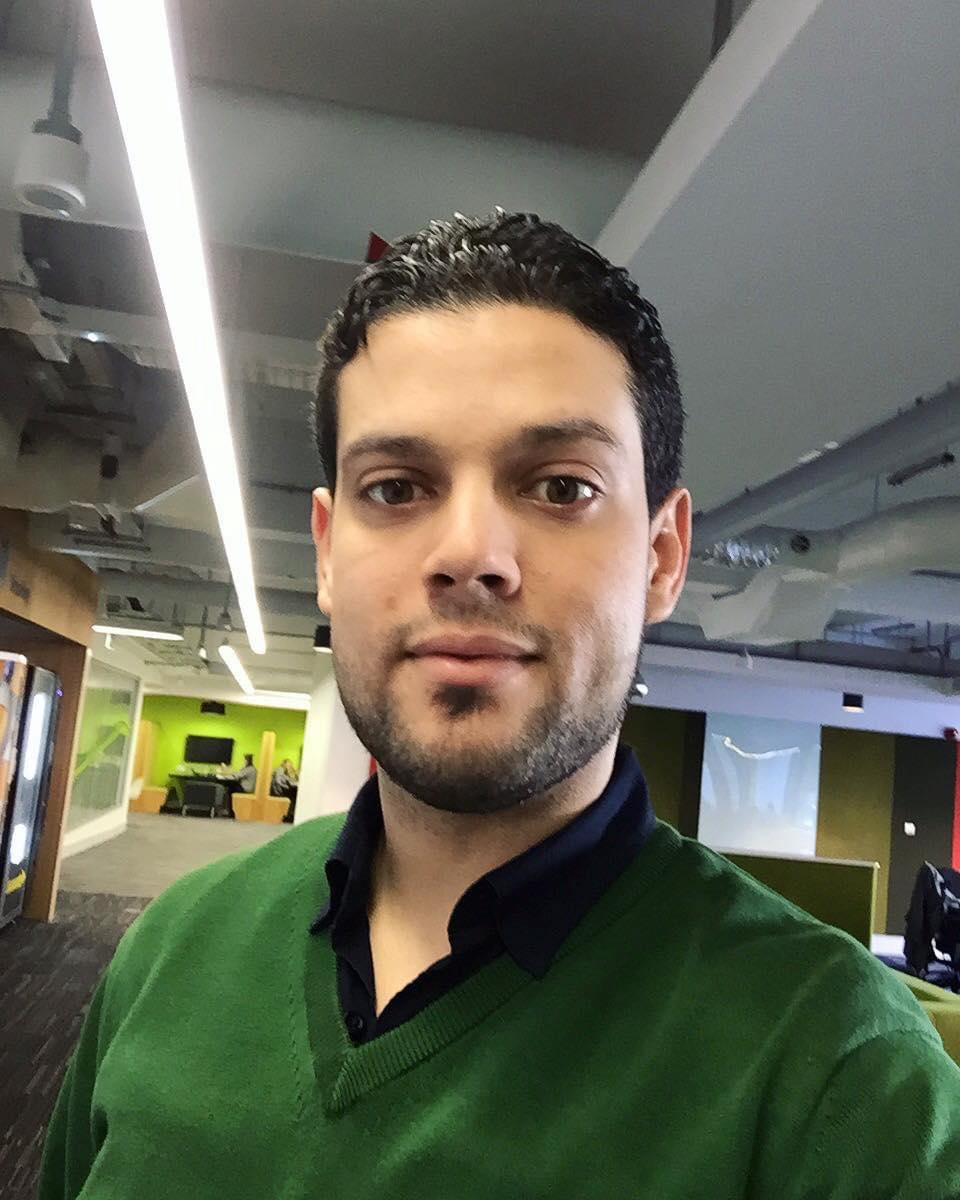}}]{Abdelaziz Salama} received his B.Sc. degree in Electrical and Electronic Engineering from Tripoli University, Tripoli, Libya, in 2009, and his M.Sc. degree in Communication, Control, and Digital Signal Processing from the University of Strathclyde, Glasgow, UK, in 2017. He is currently pursuing a PhD degree at the University of Leeds, Leeds, UK, with research interests primarily focused on Federated Learning, Autonomous Systems, and Sensing. He brings a wealth of industry experience, having worked for nine years in various roles within the fields of Telecommunication Engineering, Information Technology, and Management, both locally and internationally.
\end{IEEEbiography}

\begin{IEEEbiography}[{\includegraphics[width=1in,height=1.15in,clip]{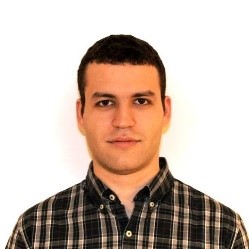}}]{Achilleas Stergioulis} received his MEng (5-year Engineering Diploma) in Electrical and Computer Engineering from Aristotle University of Thessaloniki, Greece in 2020. As part of his undergraduate diploma dissertation, he carried out research on causal analysis and inference in the presence of missing data. He completed his Master of Science (MSc) in the field of Embedded Systems Engineering from the School of Electrical and Electronic Engineering at the University of Leeds in 2021, where he focused on applying decentralised federated learning in aid of communication systems. He is currently working as a digital design engineer in the semiconductor industry.\end{IEEEbiography}

\begin{IEEEbiography}[{\includegraphics[width=1.1in,height=2in,clip,keepaspectratio]{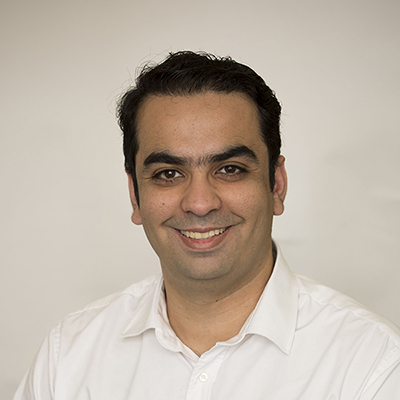}}]{Syed Ali Raza Zaidi} (M'09) is an Associate Professor at the University of Leeds in the broad area of Communication and Sensing for Robotics and Autonomous Systems. Earlier from 2013-2015, he was associated with the SPCOM research group working on US ARL funded project in the area of Network Science. From 2011-2013, he was associated with the International University of Rabat working as a Research Associate. He was also a visiting research scientist at Qatar Innovations and Mobility Centre from October- December 2013 working on QNRF funded project QSON. He completed his Doctoral Degree at the School of Electronic and Electrical Engineering. He was awarded the G. W. and F. W. Carter Prize for best thesis and best research paper. He has published 90+ papers in leading IEEE conferences and journals. From 2014-2015, he was the editor for IEEE Communication Letters and also lead guest editor for IET Signal Processing Journal's Special Issue on Signal Processing for Large Scale 5G Wireless Networks. He is also an editor for IET Access, Front haul and Backhaul books. Currently, he is serving as an Associate Technical Editor for IEEE Communication Magazine. He has been awarded COST IC0902, Royal Academy of Engineering, EPSRC, Horizon EU and DAAD grants to promote his research outputs. His current research interests are at the intersection ICT, applied mathematics, mobile computing and embedded systems implementation. Specifically, his current research is geared towards: (i) design and implementation of communication protocols to enable various applications (rehabilitation, healthcare, manufacturing, surveillance) of future RAS; and (ii) design, implementation and control of RAS for enabling future wireless networks (for e.g. autonomous deployment, management and repair of future cellular networks). \end{IEEEbiography}

\begin{IEEEbiography}[{\includegraphics[width=1.05in,height=1.2in,clip]{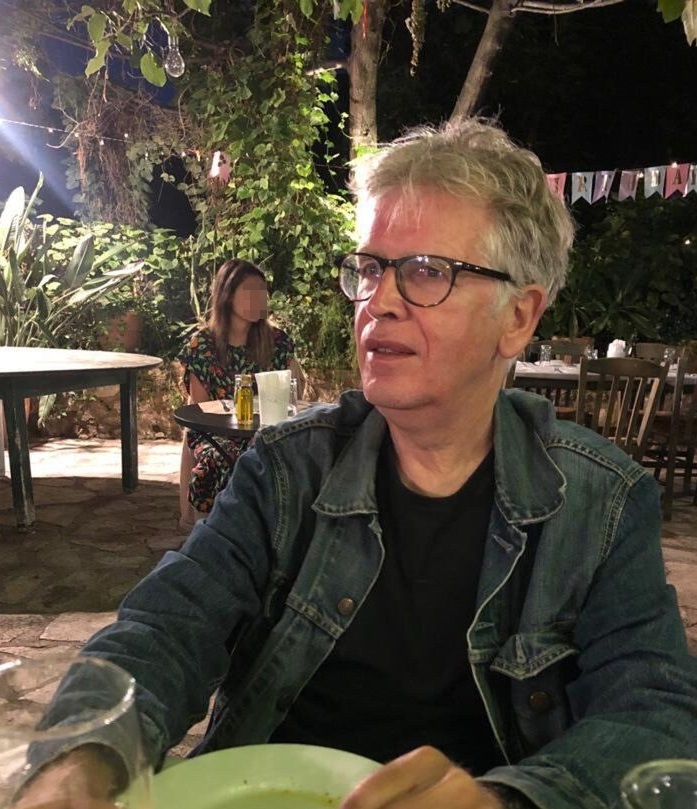}}]{Des McLernon} received both his B.Sc in electronic and electrical engineering and his MSc in electronics from the Queen's University of Belfast, N. Ireland. After working on radar systems research with Ferranti Ltd in Edinburgh, Scotland, he then joined Imperial College, University of London, UK, where he took his PhD in signal processing. He is currently a Reader in Signal Processing at the University of Leeds, UK. His research interests are broadly within the domain of signal processing for wireless communications, in which field he has around 350 research publications and also supervised over 50 PhD students. Finally, in the little spare time that remains, he plays jazz piano in restaurants and bars.\end{IEEEbiography}

\EOD

\end{document}